\documentclass[prd,preprint,a4paper,superscriptaddress,nofootinbib,11pt]{revtex4}
\usepackage{amsmath,amsfonts,amssymb}
\usepackage{txfonts}
\usepackage[T1]{fontenc}
\usepackage[safe]{textcomp}

\usepackage{csquotes}
\newcommand{\x}{x^{\mu}}
\newcommand{\p}{p^{\mu}}

\newtheorem{theorem}{Theorem}[section]
\newenvironment{proof}[1][Proof]{\begin{trivlist}
\item[\hskip \labelsep {\bfseries #1}]}{\end{trivlist}}

\begin{document}

\begin{center}
{\bf  \Large MICZ Kepler Systems in Noncommutative Space and Duality of Force Laws}
 
 \bigskip

\bigskip
Partha Guha
{\footnote{e-mail:partha@bose.res.in}}, \\  
S.N. Bose National Centre for Basic Sciences
JD Block, Sector III, Salt Lake
Kolkata - 700098, India
 \\[3mm]
E. Harikumar  {\footnote{e-mail: harisp@uohyd.ernet.in }} and Zuhair N. S. {\footnote{e-mail:zuhairns@gmail.com}}\\
School of Physics, University of Hyderabad,
Central University P O, Hyderabad-500046,
India \\[3mm] 

\end{center}
\setcounter{page}{1}
\bigskip

\begin{center}
{\bf   Abstract}
\end{center}

 In this paper, we analyze the modification of integrable models in the $\kappa$-deformed space-time. 
We show that two dimensional isotropic oscillator problem, Kepler problem and MICZ-Kepler problem 
in $\kappa$-deformed space-time admit integrals of motion as in the commutative space. We also show 
that the duality equivalence between $\kappa$-deformed Kepler problem and $\kappa$-deformed two-dimensional 
isotropic oscillator explicitly, by deriving Bohlin-Sundman transformation which maps these two systems. These 
results are valid to all orders the the deformation parameter.

 \newpage
 
 \section{Introduction}

 Interplay of quantum mechanics and general relativity is expected to bring in a fundamental length scale in 
the description of microscopic theories of gravity \cite{dop}. The existence of such a length scale is common 
to different approaches to quantum gravity. Non-commutative space-time offers a possible way to introduce this 
fundamental length scale\cite{dop,jack,Riv1,Riv2,Riv3}.
Concept of non-commutativity was first suggested by Heisenberg as a possible way to handle the UV-divergence in
 quantum theory, which was taken up by Snyder\cite{snyder} in his attempt to introduce a Lorentz invariant discrete
  space-time. 
  Non-commutative geometry is a feasible approach to probe the quantum nature of space-time at 
  Planck scale\cite{connes}. The major reason is that the underlying spacetime structure when quantum
gravity effects are included could be more complex and
may induce nonlocal effects \cite{dop}. Noncommutative field theories are inherently nonlocal and thus provides a 
framework to capture these effects. It is known that H.Yukawa \cite{Yukawa} proposed a nonlocal field theory, the 
bilocal field theory, in the same period as Snyder's noncommutative spacetime theory appeared. His motivation came 
from a unified description of elementary particles and to get divergence-free field
theories by introducing a fundamental length in space-time. Recent interest in non-commutative geometry arise in the context of string theory\cite{sw} and quantum gravity\cite{connes}.
Moyal space-time is one of the simplest example for a non-commutative space-time whose coordinates satisfy 
the commutation relations $[\hat{x}_\mu, \hat{x}_\nu]=\Theta_{\mu \nu},  $ where $\Theta_{\mu \nu}$ is a constant.
Generalisation of integrable models to non-commutative space-time and their analysis is being attempted 
in recent times. The correction to Newton's second law in Moyal's space-time and the 
resulting breakdown of rotational symmetry in a central potential has been demonstrated in \cite{rom1,rom2}. $\kappa$-deformed space-time 
is an example for a 
Lie algebra type non-commutative space-time. This space-time is defined such that the coordinates satisfy 
\begin{eqnarray}
\{ \hat{x}^i, \hat{x}^j\} = 0 \label{kappa1} \\  
\{ {\hat x}^0, {\hat x}^j \} = ax^i \label{kappa2}
\end{eqnarray}
Note that the deformation parameter \enquote*{$a$} 
has dimension of length and in the limit $a\rightarrow 0 $, we get back the commutative space-time. 
The modification to Newton's equations of motion was analysed in $\kappa$-space-time and the corrections to 
equation of motion for 
Kepler potential\cite{hari1}. Modification of Maxwell's equations and Lorentz force equation in $\kappa$-deformed 
space-time are studied in\cite{hari2,em}. The $\kappa$-deformed geodesic equation is obtained in\cite{hari3}.

Various integrable systems like Kepler problem, harmonic oscillator problem, Euler's top and Jacobi's geodesic 
flow on an ellipsoid have attracted wide attention and these systems were analyzed using the Hamilton-Jacobi method. 
The subject of integrability was studied extensively using Hamilton-Jacobi method and put on a firm foundation\cite{integ}.  
Of these integrable systems, Kepler potential and harmonic oscillator potential always have a special status owing 
to their wide applicability in physics. These systems were shown to be related to each other by a duality transformation, 
this equivalence was obtained by expressing these potentials using complex variables \cite{bohlin}. 
Well known result for the central potentials that give rise to bounded closed orbits have been generalised to 
the general class of Hamiltonians with either a Kepler-type or a harmonic oscillator-type potential\cite{iwai1,iwai2}.
Investigation to analyse the notion of integrability beyond classical systems are pursued in recent 
times\cite{besse1,besse2,palis,dnf,balles,balles2,balles3,balles4}. Generalisation of the concept of integrability 
to the relativistic domain was initiated by the study of Bertrand space-time which serves as a platform for the 
generalisation of Bertrand's theorem\cite{perlick}. It is of interest to see how these notions are modified in 
the non-commutative space-time. The study of noncommutative mechanics in a general setting could be very complicated. Therefore, in
the present work, we restrict our attention to Kepler problem in the  $\kappa$-deformed space-time.

There are few deformations of oscillator and Kepler systems, which preserve part of hidden symmetries, e.g.,
anisotropic oscillator, Kepler system with additional linear potential, two-center Kepler system, as well as their 
``MICZ-extensions''. The latter extension that has attracted attention in recent times is called the MICZ-Kepler 
system, which is a natural generalisation of a Kepler potential with an additional monopole field and hence shares 
characteristic features of Kepler potential, like the existence of generalised Laplace-Runge-Lenz vector. MICZ-Kepler 
problem belongs to the class of generalised Kepler-type potential and were investigated independently by 
McIntosh-Cisneros \cite{mc} and Zwanziger \cite{zw} in different contexts. Recently Meng \cite{Meng} formulated the 
MICZ-Kepler system as a system 
on what he called a ``Sternberg phase space''. In fact Meng \cite{Meng1} showed that the MICZ-Kepler problem also 
exists in higher dimensions, just as the Kepler problem does, where they remain superintegrable.

Different generalisations of MICZ-Kepler problem have been studied due to their important role in the investigations of hidden symmetries. Physically, MICZ-Kepler can be thought of as 
a Hydrogen atom with an additional monopole field. The problem of hydrogen atom in the presence of constant electric 
field is well known \cite{land}. An analysis of same situation with the MICZ-Kepler potential yields a linear 
Stark effect which completely lifts the degeneracy on the azimuthal quantum number $m$\cite{lev}. Presence of 
non-commutativity yields a notable non-trivial perturbative corrections to the Stark  effect in case of 
MICZ-Kepler problem \cite{den} which is not present in case of standard Hydrogen atom model\cite{cha}. Interestingly, it has also been shown that a massless particle in a non-commutative space under a harmonic oscillator field with an additional momentum-dependent potential will result in dual model of MICZ-Kepler system\cite{hor}.

In this paper, we construct some well known classical integrable models in $\kappa$-deformed space-time and study the 
consequence of the non-commutativity on the integrability of these systems.  Organisation of this paper is as follows. In the first section, we gave an overview of the current understanding on the subject. In the second section, 
we briefly state the well known results on integrability of Kepler and harmonic oscillator potential in 
commutative space-time. We also provide a brief review of representation of coordinates in $\kappa$-space-time in 
terms of ordinary commutative variables.The following subsection deals with the construction of central force problems 
in $\kappa$ space-time. We obtained the $\kappa$-deformed Hamiltonian in terms of classical variables of the systems 
of our interest and construct the corresponding integrals of motion. The integrability of central potential is studied 
with particular emphasis on the Kepler and harmonic oscillator potential. We analyze the Kepler problem and the 
isotropic oscillator problem in the $\kappa$-deformed space-time, obtaining the equations of motion and conserved 
quantities. In section 3, the duality mapping between these potential in the commutative situation is generalised 
to the non-commutative situation. We start from the $\kappa$-deformed harmonic oscillator in two dimensional space and 
show the duality equivalence between $\kappa$-deformed Kepler problem and the $\kappa$-deformed isotropic oscillator.
In section 4, we have investigated the integrability of MICZ-Kepler problem in $\kappa$ space-time. In the limit 
$a\rightarrow 0$ we retrieve the known commutative results. We conclude in section 5. Details of the duality transformation is given in the appendix.    

\section{Central force problem in $\kappa$-space-time}
In this section we review the familiar understanding of central potentials, in particular the harmonic 
oscillator system and Kepler system in the commutative space-time. 

The integrability of the harmonic oscillator and Kepler potential is responsible for the fact that 
they are the only two central potentials with closed and bounded orbits which is the content of 
Bertrand's Theorem. For Kepler system in 3-dimensional space, energy, angular momentum (3 components) and 
the Runge-Lenz vector (1 independent component) are the integrals of motion. Altogether, this system 
do have five integrals of motion. They allow for the maximal
number of functionally independent constants of motion, and at the quantum
level, these properties entails a fact: energy levels depend on a single
quantum number. A good account of treatment of Laplace-Runge-Lenz vector in non-commutative quantum mechanics can seen in\cite{pet}. Related to superintegrability, the Kepler potential displays separability
for Hamilton-Jacobi equations in several coordinate systems: this has been
called multiseparability or superseparability. 

The Laplace-Runge-Lenz vector which is a constant of motion in the Kepler problem is given by
\begin{equation}
 \vec{A}= \vec{p}\times \vec{r}-mC\frac{\vec{r}}{r}
\end{equation}
where C is a constant. Now let us consider an isotropic Harmonic Oscillator system in two dimensions. It has three integrals of motion 
known as Fradkin-Hill tensor. Fradkin-Hill tensor T whose is given by
\begin{equation}
  T_{ij}=\frac{1}{2m}p_{i}p_{j}+\frac{k}{2}x_{i}x_{j} \label{44}
\end{equation}   
where $k$ is a constant. 

Thus it is clear that both of the systems are superintegrable. Further, it is well known that there exists a duality between these two potentials. In order to facilitate this duality mapping, it is convenient to work in the complex coordinate system.

A word about convention: Throughout this paper, we will be using coordinate \enquote*{$z$}  for the harmonic oscillator system and \enquote*{$w$} for the Kepler system. Further, since our calculations are classical in nature all references to the term `commuting' would mean classical phase space variables obeying the usual Poisson bracket relation and  `non-commuting' would mean variables obeying the Poisson bracket 
relations (\ref{xx}), (\ref{pp}) and (\ref{xp}). 

\subsection{$\kappa$-space-time}   

The approach we take here in generalising the Kepler problem and isotropic harmonic oscillator to $\kappa$-deformed space-time is to first re-express the non-commutative Hamiltonian in terms of the classical variables. This allows us to map the problem in the $\kappa$-deformed space-time to an equivalent problem in commutative space-time. This map allows us to write any function of non-commutative coordinate in terms of coordinates and momenta of the commutative space as a perturbation series in powers of deformation parameter $a$. This approach has been used in deriving and analysing the $\kappa$-deformed electrodynamics\cite{em}, geodesic equation in $\kappa$ space-time\cite{hari3}, effect of $\kappa$-deformation on black hole entropy\cite{gupta}. In this approach, one first introduce a realisation of $\kappa$-deformed coordinate $\hat{x}$ and the corresponding momentum $\hat{p}$ in terms of commuting variables\cite{em, mel, mel2,mel3,mel4,mel5} as eqn. (\ref{cor}) and (\ref{mom}) . One can readily verify 
that this 
realisation will satisfy (\ref{kappa1}) and (\ref{kappa2}) with the choice $a^\mu = (a, \vec{0})$.  
\begin{eqnarray}
\hat{x}^\mu &=& x^{\mu}+\alpha x^{\mu}(a.p)+\beta(a.x)p^{\mu}+\gamma a^{\mu}(x.p) \label{cor}\\
    \hat{p}^\mu &=& p^{\mu}+(\alpha+\beta)(a.p)p^{\mu}+\gamma a^{\mu}(p.p) \label{mom} 
\end{eqnarray}
 where $\mu$ takes values $0,1,2,3$. These $\hat{x}^\mu$ and $\hat{p}^\mu$ satisfy the following relations expressed with poisson brackets and describes the fundamental structure of the $\kappa$-deformed space-time.
 \begin{eqnarray}
  \{\hat{x}^\mu,\hat{x}^\nu\}&=& a^\mu \hat{x}^\nu-a^\nu\hat{x}^\mu, \label{xx}\\  
  \{\hat{p}^\mu,\hat{p}^\nu \}&=&0, \label{pp} \\
  \{\hat{p}^\mu,\hat{x}^\nu\}&=&\eta^{\mu \nu} (1+s(a.p)+ (s+2)a^\mu p^\nu + (s+1)a^\nu p^\mu,\label{xp}
 \end{eqnarray}
 where $s=2\alpha+\beta$. $a^\mu $ is the deformation parameter, $\alpha$, $\beta$ and $\gamma$ are constants with 
 \begin{equation*}
 \gamma-\alpha=1,   \beta \in \mathbb{R}
 \end{equation*}
 
 Once we choose, $a^\mu=(a,0)$, it is clear that the above expressions reduce to those in commutative space-time as $a \rightarrow 0$.  It should be kept in mind that all the calculations in this paper are classical in nature and we are not going to discuss any quantum mechanical effects. Also, note that $\hat{x}^\mu $ and $\hat{p}^\mu $ are variables defined in terms of ordinary commuting variables.

\subsection{Dynamics in $\kappa$ space-time}

The notion of integrability is related to the existence of first integrals of motion. There are 
various methods of finding first integrals, some of them are based on symmetry methods, 
some of them relies on a direct search for invariants and some other related to Painlev\'e analysis. One of the most aesthetically appealing problem in physics is the central field problem. In 1873, J. Bertrand 
\cite{Bertrand} established that, among all central potentials, the Kepler ( $V =- 1/r$) and Hooke ($V = r^2$) problems
are very special in the sense that they are the only ones for which the bounded planar orbits are closed
for any initial conditions and of elliptic type.

The Hooke and Kepler systems are connected by a duality map as shown by Bohlin in 1911 \cite{Bohlin}. 
The complex transformation $z$ to $z^2$ takes trajectories of Hooke's Law $w^{\prime \prime} = -Cw$
to trajectories of Kepler's Law $\frac{d^2Z}{d\tau^2} = -{\tilde C} \frac{Z}{Z^3}$, where $Z(\tau(t)) = [w(t)]^2$
 and ${\tilde C} = 2(|w(0)^{\prime}|^2 + C|w(0)|^2)$.
This transformation
is called Bohlin-Sundman transformation, since Sundaman tried to find a transformation that would remove
the singularity present at the origin. This relation can be
formulated in terms of conformal mapping which is the heart of the Levi-Civita regularization scheme \cite{Levi}. 
This duality map has been rediscovered and generalized by Arnold and Vassiliev \cite{Arnold, Arnold1}.

\bigskip
We start with Hamiltonian of the form given by
\begin{equation}
  \hat{H} = \hat{T} + V(\hat{x}^{\mu}).
\end{equation}
The hat in the above equation refer to functions of variables in $\kappa$-deformed space-time. 
Using the realization of $\hat{x}^\mu$ given in eqn. (\ref{cor}), we expand any function of $\hat{x}^\mu$. 
In particular, we expand the potential $U(\hat{x}^\mu)$ and the resulting potential in commuting space variables 
is expressed as $V(x)$.  
 \begin{equation}
 U(\hat{x}^\mu)= U(\alpha x^\mu (a.p)+\beta (a.x)p^\mu+\gamma a^\mu (x.p)\label{8}
 \end{equation}
 We choose $a^\mu  = (a,\vec{0})$. 
 
 \subsection{Kinetic term and the correction to mass term}
 
 We now turn our attention to the kinetic energy term of the Hamiltonian in $\kappa$-deformed space-time expressed in terms of commuting variables. We have the expression for $\hat{p}_0$ expressed in terms of commuting variables given by 
\begin{equation}
\hat{p}_{0} =p_{0} +a (\alpha +\beta) {p_{0}}^2 + a\gamma m^{2}  \label{energy}
\end{equation}
 Sqauring the expression, we obtain
 \begin{equation}
  \hat{p}_{0}^2= p_{0}^2 +2a (\alpha +\beta) p_{0}^3 +2a\gamma m^2 p_{0} + 2\gamma (\alpha +\beta) a^2 p_{0}^2 m^2 + (\alpha +\beta) ^2 a^2 p_{0}^4 + \gamma^2 a^2 m^4  \label{p}
\end{equation}
Expressing $\hat{p}_{0}$ as a series expansion in $a$,
\begin{equation}
 \hat{p}_{0}\equiv \hat{E} = E_{0}+aE_{1}+a^2 E_{2} +...  
\end{equation}
Squaring and comparing with (\ref{p}) we obtain 
\begin{eqnarray}
p_{0}&=&E_{0}  \\
E_{1} &=& (\alpha +\beta) p_{0}^2 + \gamma m^2  
\end{eqnarray}
Note that the above result is exact as all the higher order 
corrections are identically zero. 
In the non-relativistic limit, 
\begin{equation}
  E_{0} =  \frac{p^2}{2m} + m 
\end{equation}
Then,
\begin{eqnarray}
  \hat{E} &=& \frac{p^2}{2m} + m  + a (\alpha +\beta) p_{0}^2 + a\gamma m^2  \\
    &=& \frac{p^2}{2m} + m  + a (\alpha +\beta) (p^2 + m^2)+ a\gamma m^2 \\
  &=& \frac{p^2}{2m} (1+ 2a(\alpha +\beta) m ) + m + a m^2 (\alpha +\beta + \gamma) \\
&=& \frac{p^2}{2\tilde{m}} + m + a (2\alpha+\beta +1)m^2  \label{kinetic}
\end{eqnarray}
where the relation $\gamma-\alpha =1$ is used and $\tilde{m}$ is given by 
\begin{eqnarray}
  \tilde{m} = \frac{m}{1+2a(\alpha +\beta) m}\label{modmass}
\end{eqnarray}
Note that the deformed mass includes the correction valid to all orders in the deformation parameter $a$ 
and the additional term in (\ref{kinetic}) can be neglected as it is just a shift in energy level. Interestingly, 
such a correction to the rest mass would give rise to a shift in the energy spectrum as the one seen in the case of 
the $k$-deformed Hydrogen atom \cite{harisiva}.

\subsection{Functions in $\kappa$ space-time}

Consider a function $F(\hat{x})$ in $\kappa$ space-time and express it in terms of variables in commuting space-time using the mapping 
\begin{equation}
  \hat{x_{i}}=x_{i}+\alpha x_{i}(a.p)+\beta(a.x)p_{i} \label{23}
\end{equation}
resulting from Eqn. (\ref{cor}). Using this, we express an arbitrary function of the non-commutative coordinates as  
\begin{eqnarray}
  F(\hat{x}) &=& F(\hat{x_{i}}=x_{i}+\alpha x^{i}(a.p)+\beta(a.x)p^{i}) \\
  &=& F(x) + a\alpha E_{0} \frac{dF}{dx} + \frac{(a\alpha E_{0})^2}{2!} \frac{d^2F}{dx^2} + \rm~higher~order~terms~in~a~
  \end{eqnarray}
where we have identified $p_0$ with $E_0$.  One could easily see from the above expression that the function $F(\hat{x})$ can be rewritten as a function of $x_i$ (the commutative coordinates)  and other terms which depend on the deformation parameter, $a$. Hence, we write
\begin{equation}
  F(\hat{x}) = f(x,p,a) \label{F}
\end{equation}
For power potentials we get further simplification. To see this, we start with a generic power potential in $\kappa$-deformed space-time
\begin{eqnarray}
V(\hat{x}) = C\hat{x}^n
\end{eqnarray} 
where n is a rational number. We can rewrite $V(\hat{x})$, using eqn.(\ref{23}) and identifying $p_{0}$  with $E_{0}$. In order to avoid the potential to have explicit time dependence, we set $\beta=0$
in Eqn. (\ref{23}) and thus find
\begin{eqnarray}
  V(\hat{x}) &=& C (x(1+a\alpha {E_{0}}))^n \\
   &=& \tilde{C} U(r)
\end{eqnarray}
with 
\begin{eqnarray}
\tilde{C} &=& C (1+a\alpha {E_{0}})^n \label{C} \\
U(r) &=& x^n
  \end{eqnarray}
Thus we see that any power potential in $\kappa$ space-time can be expressed in terms of a power potential in 
commuting variables, but with the coffecient of the potential depending on the deformation parameter. This inturn 
completely takes into account the effect of non-commutativity on the dynamics of the system, to all 
orders in the deformation parameter.

\section{Kepler potential}

  Now let us consider the Kepler potential with $V(\hat{x}^\mu ) = -\frac{C}{\hat{r}}$ where $\hat{r}=\sqrt{\hat{x^\mu}\hat{x_{\mu}}}$. We start with the Hamiltonian 
\begin{eqnarray}
  H = \frac{p^2}{2\tilde{m}}+V(\hat{r})\label{keplerH}
\end{eqnarray}
\begin{equation}
 \hat{r}\equiv (\hat{x}_{i}^2)^{\frac{1}{2}} = r(1+a\alpha E_{0}) \label{hatr} 
\end{equation}
Note that we have kept the non-commutative corrections valid to all orders in a.
Using this we find
\begin{eqnarray}
 V(\hat{r}) = -\frac{C}{r(1+a\alpha E_{0})} .
\end{eqnarray}
Substituting above potential in the Hamiltonian in Eqn.(\ref{keplerH}) we derive equations of motions using Poisson bracket relations. Thus we get
\begin{eqnarray}
 \dot{p}_{j} = \{p_{j}, H \} = \{p_{j}, V(\hat{r}) \} = - \frac{\partial V(x)}{\partial x_{j}}  \\
 \dot{p}_{j} = -\frac{x_{j}}{r^3} \frac{C}{(1+a\alpha E_{0})}
\end{eqnarray}
Similarly we get
\begin{eqnarray}
  \dot{x}_{j} =  \{x_{j}, H \} = \frac{p_{j}}{\tilde{m}} 
\end{eqnarray}
\begin{eqnarray}
  \overset{..}{x_{j}} = \{\dot{x}_{j}, H\} = \{\frac{p_{j}}{\tilde{m}}, H \}= -\frac{x_j}{\tilde{m}r^3} \frac{C}{(1+a\alpha E_0)} 
\end{eqnarray}
which we re-express as
  \begin{equation}
   \overset{..}{x_{j}}= -\frac{\tilde{C}}{\tilde{m}} \frac{x_{j}}{r^3}    \label{40}
  \end{equation}
  where $\tilde{C}$ is given by
  \begin{equation}
    \tilde{C} =C \frac{1}{(1+a\alpha E_{0})} 
  \end{equation}
  
  \subsection{Conservation of Angular momentum}
  In this section, we will show that the angular momentum is a conserved quantity for the Kepler problem in $\kappa$ space-time. We start with the angular momentum in the $k$-deformed space-time 
  \begin{eqnarray}
    \vec{L} = \tilde{m} \vec{r} \times \dot{\vec{r}}  \\
     \end{eqnarray}
To show that the above defined angular momentum in $\kappa$ space-time is conserved, we take the time derivative of $L$, i.e,  $\frac{d\vec{L}}{dt} = \{ \vec{L}, H\}$  which in the component  form is 
     \begin{eqnarray}
  \frac{dL_{i}}{dt} &=& \tilde{m} \{\varepsilon_{ijk} x_{j} \dot{x_{k}}, H \} \\
     &=& \tilde{m} \varepsilon_{ijk} [\{ x_{j}, H\} \dot{x_{k}} + \{ \dot{x_{k}}, H\} x_{k} ] \label{ldot}
     \end{eqnarray}
     From eqn.(\ref{40}). we note that $\overset{..}{x_{j}} \propto x_{j}$. Using this fact and also noting that $\{x_{j}, H\} = \dot{x_{j}}$, it is easy to see from eqn.(\ref{ldot}) that 
     \begin{equation}
       \dot{\vec{L}} = 0
     \end{equation}

     \subsection{Conservation of Laplace-Runge-Lenz vector} 
     
     We generalise the Laplace-Runge-Lenz vector to $\kappa$ space-time as  
     \begin{eqnarray}
        A_{i} &=& \tilde{m} \varepsilon_{ijk} \dot{x_{j}} L_{k} - \tilde{m} \tilde{C} \frac{x_{i}}{r}    
     \end{eqnarray}
      Taking the time derivative 
\begin{equation}
 \{ \vec{A}_i, H\} =  \tilde{m} \varepsilon_{ijk} \{ \dot{x_{j}}, H\} L_{k} - \tilde{m} \tilde{C} \left[\frac{\{x_{i}, H\}}{r} + x_{i} \{\frac{1}{r}, H\} \right]   \label{51}
                \end{equation}
where we have used the fact that the angular momentum is conserved in setting one term to zero. Note that
    \begin{eqnarray}
       \{ \dot{x_{j}}, H\} &=& -\tilde{C}\tilde{m} \frac{x_{j}}{r^3} \\
                \{ x_{j}, H\} &=& \dot{x_{j}} \\
          \{ \frac{1}{r}, H\} &=& \frac{-x_{j}\dot{x_{j}}}{r^3}
    \end{eqnarray} 
   where in obtaining the last term we have used the result
    \begin{equation}
     \{f(q), p\} = \frac{\partial f}{\partial q} .
    \end{equation}
    Using these along with the expression
    \begin{equation}
      \varepsilon_{ijk} x_{j} L_{k} = \tilde{m} (x_{i} x_{j} \dot{x_{j}} - x_{j} x_{j} \dot{x_{i}} ) 
    \end{equation}
    in eqn.(\ref{51}) show the conservation of the Laplace-Runge-Lenz vector,
        \begin{equation}
       \{ \vec{A}, H\} = 0
    \end{equation}
In comparison to the expression in commutative case, we note that the $\kappa$-deformed angular momentum and Laplace-Runge-Lenz vectors have modified coefficients $\tilde{C}$ and $\tilde{m}$, and these modifications vanishes as we set the deformation parameter to be zero, reproducing the correct commutative limit. 

\section{Isotropic oscillator in two dimensions}

  In this section, we study isotropic oscillator in two dimensional deformed space. The aim is to understand how the integrals of motion get modified in the above non-commutative setting. The dimension is restricted to two for future convenience as we plan to analyze a duality between isotropic oscillator and Kepler problem in the next section.
  For $\kappa$-deformed isotropic oscillator in two dimensions, the potential is 
  \begin{eqnarray}
      V(\hat{x}^\mu )&=&\frac{k}{2} \hat{r}^2
       \end{eqnarray}
and the Hamiltonian is given by 
\begin{equation}
 H= \frac{p_{i}^2}{2\tilde{m}}+\frac{k(1+2a\alpha E_{0})^2}{2}r^2
\end{equation} 
  where $r$ is the magnitude of position vector. It should be noted that all the equations are valid to all orders in the deformation parameter.

  Following a calculation along the similar lines as in section 2.4, we obtain the equation of motion associated with the deformed isotropic oscillator as
  \begin{equation}
  \tilde{m}\overset{..}{\vec{r}}=-\tilde{k}\vec{r} \label{60}
  \end{equation}
  where 
  \begin{equation}
    \tilde{k}=k(1+2a\alpha E_{0})^2
  \end{equation}
  We obtain the first integral of motion as 
  \begin{eqnarray}
    E&=&\frac{\tilde{m}}{2} (\dot{x}^2 + \dot{y}^2) + \frac{\tilde{k}}{2}(x^2+y^2), \\
    E&=&\frac{\tilde{m}}{2} \lvert \dot{w}^2 \rvert + \frac{\tilde{k}}{2}\lvert w^2 \rvert. \label{ehar}
  \end{eqnarray}
   where the second expression is the energy expressed in terms of complex coordinate $w=x+iy$.
   Further, the equation of motion in complex coordinate system is given by
   \begin{equation}
\overset{..}{w}+\frac{\tilde{k}}{\tilde{m}}w=0 \label{64}
\end{equation}
Interestingly, there is no $\bar{w}$ dependence in the equation of motion and this fact can be exploited to obtain a conserved quantity. To see this we multiply eqn.(\ref{64}) with $\dot{w}$ followed by an integration in time. Thus we obtain  
    \begin{eqnarray}
    \mathcal{F} &=& \frac{\tilde{m}}{2}\dot{w}^2+\frac{\tilde{k}}{2}w^2 \label{fh}
   \end{eqnarray} 
   which can be re-written as 
   \begin{equation}
       \mathcal{F}=\mathcal{F}_{1}+i\mathcal{F}_{2} 
   \end{equation}
   with $\mathcal{F}_{1}=T_{xx}-T_{yy}$ and $\mathcal{F}_{2}=T_{xy}$, where $T_{ij}$ are the components of the Fradkin-Hill Tensor given in eqn. (\ref{44}), with coefficients getting $a$ dependent corrections, valid to all orders in $a$. This Fradkin-Hill tensor is analogous to the Laplace-Runge-Lenz vector of the Kepler problem. The physical significance of this tensor can be easily seen from its eigenvalue problem. The eigenvectors of this tensor determine the orientation of the orbit of the two dimensional harmonic oscillator potential. It is interesting to note the fact that the general solution for a two dimensional harmonic oscillator problem being an ellipse with centre at the origin requires a tensor quantity to determine the orientation of the orbit rather than a vector as in the case of Kepler problem where we have an ellipse with centre at one of the focii. The presence of such a conserved quantity can be attributed to the periodicity of the potential\cite{frad}.  
   
   Also note that $Tr(T_{ij}) = E$. Note that our equations of motion (\ref{60}) and (\ref{40}) are similar to that in commutative space-time, except that the coefficients are modified. This suggest that the Bohlin-Sundman transformation between the Isotropic Oscillator and Kepler system in commutative space-time is amenable for generalisation to the $\kappa$-deformed space-time. We analze this duality in the next section.
   
  \section{Duality Equivalence between Isotropic Oscillator and Kepler potential} 
  
We study here the duality mapping between isotropic oscillator problem in two dimensional space and the Kepler  problem. We start with the well known result of Kepler's second law to re-express the energy of the harmonic potential in terms of the complex coordinates $z$ of Kepler potential. This is used to show the conservation of Fradkin-Hill tensor as a consequence of the conservation of Laplace-Runge-Lenz vector.
  
 Now, consider the Bohlin-Sundman mapping, 
\begin{equation}
z\rightarrow w=z^2 \label{boh}
\end{equation} 
This gives a correspondence between isotropic oscillator and Kepler potential. This mapping can be used to exhibit the dual relation between the above systems. In order to show this, take $t,\tau$ to be the time parameter for the isotropic oscillator and Kepler system, respectively. Kepler's second law (constancy of areal velocity) implies $\frac{d\tau}{dt} = \lvert w\rvert ^2 $\cite{grandati}. Using Bohlin-Sundman map and the relation between the time parameters, we obtain an expression between the time derivatives of Kepler problem and harmonic oscillator problem. In this way, one would be able to express the $\kappa$-deformed Hamiltonian of harmonic oscillator in terms of the coordinates of Kepler problem, i.e, 
   \begin{eqnarray}
    w&=&\sqrt{z} \Rightarrow \lvert w \rvert ^2 = \lvert z \rvert, \\
    \dot{w}&=&\frac{1}{2}\dot{z} \sqrt{z}^{-1}, \\
    \dot{w}&=&\frac{1}{2}z'\lvert z\rvert \sqrt{z}^{-1}.   \label{30}
   \end{eqnarray}
   where a dot ($^.$) denotes differentiation with respect to $t$ and a prime ($\prime$) denotes differentiation with respect to $\tau$. We obtain from the above equation
   \begin{equation}
   \dot{w}^2=\frac{1}{4}\bar{z}z'^2.   \label{25}
   \end{equation}
Substituting eqn.(\ref{25}) in eqn.(\ref{ehar}), we obtain the energy expression for harmonic oscillator in terms of the coordinates of Kepler potential as

   \begin{eqnarray}
    E_{H.O.} &=& \frac{\tilde{m}}{8}\bar{z}z'^2+\frac{\tilde{k}}{2}\lvert z\rvert 
       \end{eqnarray}
 while the energy of the Kepler system is given by 
 \begin{equation}
 E_{K}=\frac{\tilde{m}}{2}\lvert z'\rvert ^2 - \frac{C}{\lvert z \rvert} \label{ekep}
 \end{equation}
   With Bohlin-Sundman mapping (\ref{boh}), one could view the orbit(solution) for the harmonic oscillator problem as a dual motion to the Kepler orbit. For this dual motion, we have $E=-2\tilde{k}$ (see appendix). Using eqn.(\ref{30}) in eqn,(\ref{fh}), we obtain 
     \begin{eqnarray}
         \mathcal{F}&=&\frac{\tilde{m}}{8}z'(\bar{z}z'-z\bar{z}')+\tilde{C}\frac{z}{4\lvert z\rvert} \\
       &=&\frac{1}{4}(i\tilde{L}z'+\tilde{C}\frac{z'}{\lvert z \rvert} \\
       &=&-\tilde{A}/4\tilde{m}.
     \end{eqnarray}
   It is clear that the Fradkin-Hill tensor is mapped into the Laplace-Runge-Lenz vector with the transformation of coordinates mapping $w$ to $z$ as given in eqn. (\ref{boh}). 
   
    In the above sections, we started with the Kepler and Harmonic Oscillator potential in $\kappa$ space-time and expressed the corresponding Hamiltonians and the resulting integrals of motion in terms of variables defined in terms of commutative space-time. Finally, we showed that the Bohlin-Sundman transformation maps $\kappa$-deformed harmonic oscillator to $\kappa$-deformed Kepler problem, validating equivalence between these two models in the $\kappa$-deformed space-time. 
  
      \section{MICZ-Kepler system}
    In this section, we study the modifications to the classical MICZ-Kepler system in the $\kappa$ space-time. In the commutative space-time, MICZ-Kepler system is described by the Hamiltonian 
    \begin{eqnarray}
      H&=&\frac{p^2}{2m}+\frac{\lambda^2}{2mr^2}-\frac{\mu}{r}       
    \end{eqnarray} 
    with $x_{i}$, $p_{j}$ satisfying the relations 
    \begin{eqnarray}
      [x_{i},x_{j}]_{P.B.}&=&0, \\ 
      {[p_{i},p_{j}]}_{P.B.}&=&\lambda \varepsilon_{ijk}\frac{x_{k}}{r^3},\\ 
      {[x_{i},p_{j}]}_{P.B.}&=&\delta_{ij}. 
    \end{eqnarray}
    Now we generalise the problem to $\kappa$ space-time. For this, we express the variables $\hat{\x}$,$\hat{\p}$ in terms of the variables in the commutative space-time. Using eqns. (\ref{cor})and (\ref{mom}) and setting $\beta =0$, we get 
    \begin{eqnarray}
    \hat{x_{i}}&=& x_{i}(1+a\alpha E), \\
     \hat{p_{i}}&=&p_{i}(1+a\alpha E). 
    \end{eqnarray}
   where the $x_{i}$ and $p_{i}$ satisfy the the above Poisson bracket relations and E is the energy of system when $a=0$ (see: eqn.(\ref{energy})).
    
          We then obtain the Hamiltonian for the MICZ-Kepler system in $\kappa$ space-time as 
      \begin{equation}
        H=\frac{\hat{p}^2}{2m}+\frac{\lambda^2}{2m\hat{r}^2}-\frac{\mu}{\hat{r}} \label{hammic} 
      \end{equation}
     Next we express the Hamiltonian in terms of commuting variables, using the map from  $\kappa$ space-time to commutative space-time, 
     \begin{eqnarray}
 \frac{1}{\hat{r}}&=&\frac{1}{r(1+a\alpha E_{0})}, \\
 \frac{1}{\hat{r}^2}&=&\frac{1}{r^2(1+2a\alpha E_{0})^2}
    \end{eqnarray}
    where the above expressions are valid to all orders 
    in the deformation parameter, '$a$'.
    
    Using these equation we re-express the Hamiltonian in eqn.(\ref{hammic}) as 
    \begin{eqnarray}
      H&=&\frac{p^2}{2\tilde{m}}+\frac{\lambda^2}{2\tilde{m}r^2}\frac{1}{(1+2a\alpha E_{0})^2}-\frac{\mu}{r}\frac{1}{(1+a\alpha E_{0})}  \label{ham}
    \end{eqnarray}
        Thus we see that the Hamiltonian in eqn.(\ref{ham}) is exactly similar in form as the commutative space-time except for the presence of $a$ dependent factors which brings in the non-commutative features into the dynamics.  In addition, note that the modification to the coefficients goes to zero as we let the deformation parameter $a$ to zero, thereby leading to expected commutative limit.
     
    We define
    
    \begin{eqnarray}
      \tilde{\lambda} = \frac{\lambda}{1+2a\alpha E_{0}} \\
      \tilde{\mu} = \frac{\mu}{1+2a\alpha E_{0}}
    \end{eqnarray}

    The equation of motion following from the Hamiltonian in eqn.(\ref{ham}) is
     \begin{equation}
      \tilde{m}\overset{..}{\vec{r}}+\frac{\tilde{\lambda}}{r^3}\frac{\vec{L}}{\tilde{m}}+\frac{\tilde{\mu}}{r^3}\vec{r}-\frac{\tilde{\lambda}^2}{2\tilde{m}r^4}\vec{r}=0 \label{mic}
    \end{equation}
    where $\vec{L}=\tilde{m}\vec{r}\times \dot{\vec{r}}$ and $\tilde{m}$ is given by eqn.(\ref{modmass}). Note that $[L_{i},H]_{P.B.}=0$, showing that $\vec{L}$ is an integral of motion, even in $\kappa$-deformed space-time. 
    Taking cross product of eqn.(\ref{mic}) with $\vec{r}$, we get
    \begin{equation}
      \tilde{m}\vec{r}\times \overset{..}{\vec{r}}+\frac{\tilde{\lambda}}{r^3}\frac{\vec{r}\times\vec{L}}{\tilde{m}}=0 \label{88}
    \end{equation} 
     Using
      \begin{eqnarray}
      \frac{d\vec{r}}{dt}&=&\dot{r}\hat{r}+r\dot{\theta} \hat{\theta}
      \end{eqnarray}
      we find the deformed angular momentum to be 
      \begin{eqnarray}
        \frac{d\vec{L}}{dt}&=&\tilde{m}\vec{r}\times \dot{\vec{r}} 
      =\tilde{m}\vec{r}\times r\dot{\theta}\hat{\theta}, \\
    {\rm and~~~~}  \vec{r}\times \vec{L}&=&-r^3 \frac{d\hat{r}}{dt},   
      \end{eqnarray}
 Using the eqn.(\ref{88}) becomes
     \begin{eqnarray}
      \tilde{m}\vec{r}\times \overset{..}{\vec{r}} -\frac{\tilde{\lambda}}{\tilde{m}} \dot{\hat{r}}&=&0
   \end{eqnarray}
   which can be expressed as
\begin{eqnarray}
      \frac{d}{dt}(\tilde{m}\vec{r}\times \dot{{\vec{r}}} -\frac{\tilde{\lambda}}{\tilde{m}} \hat{r})&=&0 
\end{eqnarray}
This allow us to define the generalised angular momentum as       
        \begin{eqnarray}
       \vec{\mathcal{L}}&\equiv& \tilde{m}\vec{r}\times \dot{{\vec{r}}} -\frac{\tilde{\lambda}}{\tilde{m}} \hat{r} =\vec{L}-\frac{\tilde{\lambda}}{\tilde{m}} \hat{r}.  \label{95}
        \end{eqnarray}
which is conserved and is an integral of motion. Further, cross product of eqn.(\ref{mic}) with $\vec{\mathcal{L}}$ can be written as
     \begin{eqnarray}
     \frac{d}{dt}(\tilde{m}\dot{\vec{r}}\times \vec{\mathcal{L}}-\tilde{\mu} \hat{r})&=&0,   
     \end{eqnarray}
     showing that 
\begin{equation}
\vec{J}=\tilde{m}\dot{\vec{r}}\times\vec{\mathcal{L}}-\tilde{\mu} \hat{r},
\end{equation}   
     is a conserved quantity and this $\vec{J}$ is the generalised Laplace-Runge-Lenz vector.
     It can be easily verified that $\vec{J}$ and $\vec{\mathcal{L}}$ satisfies the constraints 
   \begin{eqnarray}
   \vec{J}^2=2\vec{L}^2H+\tilde{\mu}^2 ; ~~\vec{J}\cdot\vec{\mathcal{L}}=\tilde{\lambda}\tilde{\mu}.
\end{eqnarray}    
Hence, Hamiltonian ($H$), generalised angular momentum ($\vec{\mathcal{L}}$) and generalised Laplace-Runge-Lenz vector 
($\vec{J}$) together constitute the five linearly independent integrals of motion of the system. Thus we show that 
the MICZ-Kepler system is indeed a superintegrable system in the $\kappa$ space-time, 
to all orders in the deformation parameter $a$.

   \section{Conclusion}

   In the $\kappa$-deformed space-time, we have constructed and investigated the Kepler problem, isotropic oscillator 
   problem in two dimension and MICZ-Kepler system. We started with the Hamiltonian defined in the $\kappa$-deformed 
   space-time which is obtained by taking the appropriate limit of the $\kappa$-deformed energy-momentum relation. 
   Using the map between the coordinates and momenta of $\kappa$ space-time to that in commutative space-time, we 
   re-express this $\kappa$-deformed Hamiltonian in terms of commuting variables. The $\kappa$-deformed Hamiltonians 
   obtained are of the same form as the commutative ones except that the coefficients modified and have \enquote*{$a$}
   dependence. These \enquote*{$a$}  dependent coefficients have smooth commutative limit. The integrability of 
   $\kappa$-deformed Kepler problem and $\kappa$-deformed oscillator are established and the corresponding integrals
   of motion were found. Further, we have demonstrated that the duality mapping do exist in the $\kappa$ space-time 
between  the Kepler and harmonic oscillator problems. We have derived the Bohlin-Sundman mapping explicitly after 
expressing  the coordinates in terms of complex variables. 

   Next, we considered the MICZ-Kepler problem which is an example of Kepler potential with an additional monopole 
   field. We utilised the expansion of non-commuting variable in terms of commuting variables to obtain the 
   $\kappa$-deformed Hamiltonian of MICZ-Kepler. We have shown the integrability of this system and explicitly 
   constructed the corresponding conserved quantities.

  Recently it has been shown that in a space whose co-ordinates obeying a Lie algebra type commutation relation,
   only Kepler potential supports closed orbits\cite{afh}. Though $\kappa$-space-time is an example of a
non-commutative space-time whose coordinates obey Lie algebraic type commutation relations, here, the non-commutativity is 
only between the time and space co-ordinates while space co-ordinates do commute among themselves (see 
Eqn.(\ref{kappa1}) and Eqn.(\ref{kappa2})). This should be contrasted the Lie algebra type relation between the 
space co-ordinates considered in\cite{afh}. The effect of the non-commutativity of the $\kappa$-deformed space-time
enters the dynamics through the corresponding Hamiltonians. But as we have shown, $\kappa$-deformation do not alter 
the form the Hamiltonian and only changes the coefficients appearing in different terms of the Hamiltonians for Kepler 
potential and harmonic oscillator considered here. These changes lead to the deformation of the integrals of motion.

   \section*{Acknowledgment}
   ZNS acknowledges the support from CSIR, India under the JRF scheme. PG wants to thank Anindya Ghose Choudhury, Guowu Meng
and Peter Leach for many enlightening discussions.
\appendix
\section{}
In this appendix, we prove the duality between Kepler potential and isotropic potential using the Bohlin map.  We also obtain as a by product, a relation between the coefficient in equation of motion of Kepler problem and isotropic oscillator problem.
\begin{theorem}
Let the motion of point \enquote*{$w$} in the complex plane be governed by the equation,
\begin{equation}
\overset{..}{w}=-\frac{\tilde{k}}{\tilde{m}}w \label{hook}
\end{equation}
 Then a point $z(\tau)=w(t)^2$, where $\frac{d\tau}{dt}=\lvert w \rvert ^2$ satisfies the equation \begin{equation}
 z^{\prime \prime}+\frac{\tilde{C}}{\tilde{m}}\frac{z}{r^3}=0
 \end{equation}
  where 
  \begin{equation}
  C=2(\lvert \dot{w}(0)\rvert ^2 +\frac{\tilde{k}}{\tilde{m}}\lvert w(0)\rvert ^2). \label{c}
  \end{equation}
\begin{proof} Using chain rule on $z=w^2$, we obtain
\begin{eqnarray}
\frac{d^2 z}{dt^2}&=& \frac{1}{\lvert w\rvert ^2} \frac{d}{dt} (\frac{1}{\lvert w\rvert ^2} \frac{dw^2}{dt}) \\
&=&\frac{2}{w\bar{w}}\frac{d}{dt}(\frac{1}{\bar{w}}\frac{dw}{dt}) \\
&=&-\frac{2}{w\bar{w}^3}\frac{dw}{dt}\frac{d\bar{w}}{dt}+\frac{2}{w\bar{w}^2}\frac{d^2w}{dt^2}
\end{eqnarray}
the second term in the last equation can be re-written using (\ref{hook}) as
\begin{equation}
\frac{d^2 z}{dt^2}= -2w^{-1}\bar{w}^{-3}[(\lvert \dot{w}\rvert ^2 +\frac{\tilde{k}}{\tilde{m}}\lvert w\rvert ^2)] 
\end{equation}
Using (\ref{ehar}), we get that the term in the square bracket is a constant. Now re-expressing $w$ in terms of $z$, we obtain the required result. 
\end{proof}
 \end{theorem}
 Thus we have established the duality between an isotropic oscillator and a Kepler problem. Interestingly, we have obtained a relation between $\tilde{C}$ and $\tilde{k}$
 given by (\ref{c}).Substituting (\ref{c}) in (\ref{ekep}) gives us after simplification
 \begin{equation}
 E_K= -2\tilde{k}
 \end{equation}
 This result is crucial for our analysis as it says that the energy for a Kepler system is related to the coupling strength $k$ of the harmonic oscillator which is its corresponding dual problem.

\bigskip

We can extend this connection to MICZ-extension.
\begin{theorem}
Suppose the motion of a point in the complex plane is given by
$w(t)$ and satisfies
\begin{equation}
\ddot{w} = Kw + \frac{Cw}{|w|^4}
\end{equation}
Then a point following the trajectory $z(\tau(t)) = w^2$, where $\frac{d\tau}{dt} = |w|^2$,
moves according to Newton's law 
\begin{equation}
\frac{d^z}{d\tau^2} = -\frac{4z}{|z|^3}({\cal E} - \frac{C}{|z|}),
\end{equation}
where $$ {\cal E} = \frac{1}{2} \big(|\dot{w}|^2 + |w|^2 + \frac{C}{|w|^2}\big) $$
is the energy of the
oscillator equation and is constant along the trajectories of motion. 
\end{theorem}

The proof of this theorem is similar to the previous one.

   \end{document}